**Are gate-all-around 2D CFETs the optimal architecture for the A2 node and beyond?**

*Fengben Xi*, Gautam Gaddemane, Anshul Gupta, Sheng Yang, Aryan Afzalian, Quentin Smets, Maarten Van de Put, Kaustuv Banerjee, Tom Schram, Devin Verreck, Ward Janssens, Juergen Boemmels, Xiangyu Wu, Thomas Chiarella, Jérôme Mitard, Gouri Sankar Kar, Geert Hellings, Cesar Javier Lockhart de la Rosa*

***Imec, Kapeldreef 75, 3001 Leuven, Belgium***
[*]Correspondence: Fengben.Xi@imec.be

**Abstract**

**As logic scaling enters the angstrom era, vertically stacked complementary field-effect transistors (CFETs) based on atomically thin two-dimensional (2D) semiconductors offer a potential route to extend device scaling beyond the A2 node. Here, we develop an A2-oriented at a CPP of 36 nm and Lg of 10 nm integration flow and present initial demonstrations of several key process modules. Despite their atomically thin channels, 2D GAA CFETs do not provide a contacted poly pitch scaling advantage over Si GAA CFETs at the A2 node, because contact formation constraints impose a similar minimum CPP of 36 nm. We also combine a critical assessment with a multiscale power–performance–area (PPA) evaluation framework spanning quantum transport simulations, compact-model generation, A2-targeted 2D CFET gate-all-around (GAA) integration-flow definition, parasitic extraction and circuit-level benchmarking. Our analysis, however, shows that the expected benefits of 2D GAA CFETs are strongly constrained by non-idealities, in particular high contact resistance and dominant layout-induced parasitic capacitances. Although architectural optimization can improve the $I_{eff}/C_{eff}$ ratio, the associated rise in absolute capacitance limits circuit-level gains. Meaningful progress will require co-optimization of contacts, transport and parasitics, together with 2D-specific CFET architectures.**

## 1.Introduction

With the rise of data-intensive workloads in artificial intelligence (AI) and high-performance computing (HPC), global computing demand is skyrocketing even as power budgets remain constrained[1,2]. Sustaining this growth requires continuous transistor scaling and new design innovations to increase performance-per-watt[3,4]. As shown in Fig. 1a, the remarkable miniaturization of silicon transistors has delivered exponential gains: planar MOSFETs gave way to FinFETs and, more recently, to gate-all-around (GAA) nanosheet FETs, pushing logic into the single-digit nanometer regime[5–10]. Today, as the industry approaches technology nodes often termed the 'angstrom era' – where critical features approach atomic-scale dimensions – classical scaling is nearing a plateau[3,5,11].

Advanced GAA transistors have demonstrated gate lengths approaching 10 nm, enabled by superior electrostatic control. For example, IBM's research alliance demonstrated stacked nanosheet FETs with a gate length of approximately 12 nm by thinning the silicon channel to about 5 nm[12]. However, scaling gate lengths significantly below ~10 nm in silicon requires sub-3 nm channel thickness, which introduces severe challenges such as increased quantum tunnelling and mobility degradation[11]. Furthermore, continued layout compaction severely limits the effective channel width ($W_{ch}$) available to each transistor, as aggressive reductions in standard-cell height inherently constrain the total width that can be allocated per device[13]. This reduction of width lowers the per-transistor drive current and exacerbates the parasitic capacitance per unit channel width. The combined effect is that simply making silicon transistors smaller no longer guarantees significant speed or energy efficiency gains in circuits.

If conventional scaling hits a wall, what could come next? One avenue under intense exploration is to adopt new channel materials and 3D integration schemes to extend CMOS performance. Two-dimensional (2D) semiconductors are at the forefront of this exploration. Atomically thin 2D materials (e.g., monolayer $MoS_2$, $WSe_2$) serve as transistor channels only a few atoms thick[14,15]. Their extreme thinness affords unprecedented electrostatic control, enabling robust switching even at gate lengths of a few nanometers[16,17]. Fig. 1b and c highlight the strong and beneficial impact of reducing channel thickness, $t$, on gate length scaling, $L_g$, by comparing the density functional theory–non-equilibrium Green's function (DFT–NEGF)-simulated scaling potential of mono- (1L, $t \approx 0.7$ nm), bi- (2L, $t \approx 1.4$ nm) and tri- (3L, $t \approx 2.1$ nm) layer $MoS_2$ and $t = 4.5$ nm Si nanosheet (NS) FETs in terms of subthreshold swing (SS) and drive performance ($I_{ON}$ at fixed $I_{OFF}$ and $V_{DD}$). By reducing $t$, electrostatic integrity (SS, Fig. 1b) is maintained down to lower $L_g$. This translates into strongly improved $I_{ON}$ (at fixed $I_{OFF}$ and $V_{DD}$, Fig. 1c) as we scale $L_g$. As a result, the 1L-$MoS_2$ offers the best scaling potential, maintaining the required transistor current target for logic operation[5] down to $L_g \approx 4$ nm, while the $t = 4.5$ nm Si NS can only scale down to $L_g \approx 12$ nm.

The hope is that 2D materials could unlock new scaling pathways, especially when combined with vertical integration. For instance, a monolithic complementary FET (CFET)—where a *p*-FET and *n*-FET share the same footprint in a vertical stack—can double transistor density without shrinking horizontal features. If 2D semiconductors are introduced as the channels in such GAA CFET structures, one might simultaneously attain ultimate gate-length scaling and extreme device density. This vision has spurred significant research: early experiments have demonstrated 2D GAA transistors with near-ideal subthreshold behavior[18] and even initial 2D CFET prototypes[19,20], reflecting worldwide interest in this direction.

A realistic assessment of 2D CFET prospects, however, must acknowledge two caveats. First, silicon is not yet obsolete. A recent demonstration by Intel showed GAA "ribbon" transistors with a Sub-10 nm physical gate length (achieved using an ultrathin ~2 nm silicon ribbon channel) – indicating that silicon devices can still be pushed to such extremes, albeit with significant process complexity[21]. Indeed, industry roadmaps (e.g., imec's in Fig. 1a) project that 2D channels, if adopted at all, would arrive only after fully exhausting silicon CFET technology—beyond the final all-silicon CFET node (beyond ~A2). In other words, 2D materials are being pursued as a potential next step after pushing silicon to its limit, not as an immediate replacement. Second, 2D CFETs face major integration and design challenges. Wafer-scale growth or transfer of high-quality monolayer films, reliably incorporating them into intricate 3D device stacks, and achieving ultra-low contact resistance without doping[22,23] remain open research problems. Moreover, even if those hurdles are overcome, we must consider whether the traditional GAA CFET architecture—meticulously optimized for silicon—will be optimal for 2D channels. A key question now is whether integrating 2D channels within an advanced GAA/CFET architecture beyond A2—the strategy currently favored in industry roadmaps—can deliver the expected boosts in performance, power, and area.

This perspective critically evaluates the potential of 2D CFETs considering these challenges. We first examine how and where 2D channels might realistically be introduced beyond the ultimate silicon CFET node (A2), by proposing a sequential CFET integration scheme. We then review the state-of-the-art in GAA nanosheet transistors and identify where scaling pressures demand new solutions compared with Si technology. Next, using these structures, we present a power–performance–area (PPA) analysis of prospective 2D GAA/CFET devices, evaluating whether they can meet the A2 iso-frequency scaling target.

## 2. Sequential 2D-CFET integration scheme for the A2 node and beyond

A process flow of 2D-channel CFET with GAA architecture at a contacted poly pitch (CPP) = 36 nm and gate length ($L_g$) = 10 nm for a target A2 technology node is shown in Fig. 2. The proposed flow employs a fully sequential CFET integration[24–26], enabling independent fabrication of the top and bottom devices. In contrast, a monolithic CFET integration[27,28] involves multiple common processes for top and bottom devices.

One of the key reasons for using sequential integration is to simplify the atomic layer deposition (ALD) deposition bottleneck of high-κ and gate work-function metal (WFM) fully around the channel encountered in a common replacement metal gate (RMG) process of monolithic CFET. At $L_g$ = 10 nm, the vertical channel-channel gap is ~5–7 nm. In a common RMG process, after the deposition of high-κ and gate WFM deposition of the first device, the vertical gap may be pinched off, providing limited or no access to carry out WFM patterning[29] and/or deposition of WFM of the second device. Sequential CFET integration alleviates this bottleneck thanks to the independent processing of top and bottom RMG modules. Other general process simplifications with sequential CFET over monolithic CFET include lower-aspect-ratio etches and depositions.

The sequential flow starts with the formation of the top NMOS device stack (Fig. 2a) comprising two 2D channels (e.g. $MoS_2$), thin protection layers (PL) that encapsulate the channels, sacrificial layers (SAC), and a bottom dielectric isolation (BDI) layer. Key fabrication steps are the nanosheet etch, dummy gate formation, inner spacer formation, SAC removal, NMOS RMG, RMG cut, source/drain (S/D) contact formation, and front-side back-end-of-line (BEOL) formation (Fig. 2b–m). After completion of the NMOS device, the wafer is bonded to a carrier wafer, flipped, and thinned from the substrate until the BDI is reached. The bottom PMOS device formation is then carried out in a similar sequence. The bottom PMOS stack is then formed on top of the BDI layer (Fig. 2n), including the PMOS 2D channels (e.g. $WSe_2$), followed by PMOS RMG integration (Fig. 2q).

A further opportunity with sequential CFET integration is to easily fabricate a gate-to-gate via (VGG). This provides a top-gate-to-bottom-gate routability option in CFETs, that is, a split-gate functionality which is a standard-cell-library-level scaling booster[30]. VGG processing is more complex in a monolithic CFET, requiring more processing steps with higher complexity. Here again, the BDI layer is utilized through VGG patterning and metallization inside the BDI layer. The BDI layer not only ensures small within-wafer height uniformity of VGG but also provides vertical edge placement error (VEPE) margin to maintain top and bottom gate isolation distance.

An important simplification in the flow is offered by using 2D transition metal dichalcogenide (TMD) nanosheets rather than Si nanosheets. We assume that the 2D-TMD RMG process does not require a high-temperature (T) reliability anneal (T > 800°C) typically used in a Si RMG process[31,32]. Similarly, other anneals at T > 400°C such as interlayer dielectric zero (ILD0) densification anneal may be avoided at the A2 node. This makes the 2D CFET front-end-of-line (FEOL) flow compatible with a typical Cu-based back-end-of-line (BEOL) stack. Hence, for 2D-CFET, we can employ a simpler, fully sequential process flow with a single wafer flip as follows: top RMG → top S/D contact → front-side BEOL → wafer flip → bottom RMG → bottom S/D contact → back-side BEOL. Note that the front-side BEOL can be exposed to bottom RMG and other processes because of their BEOL-stack compatibility. A sequential Si CFET process flow, however, would require two wafer flips to avoid BEOL exposure to high T during RMG process as follows: top RMG → wafer flip 1 → bottom RMG → bottom S/D contact → back-side BEOL → wafer flip 2 → top

S/D → front-side BEOL.

The 2D GAA flow also differs from the Si GAA flow in the S/D contact module. In the proposed 2D flow, the S/D contacts are formed by direct metallization on the exposed 2D sheets (Fig. 2k, s)[33]. This is different from the Si GAA flow, in which doped Si or SiGe is epitaxially grown selectively from the exposed Si sheets[27]. The differences between S/D contact formation in 2D versus Si, and the related contact resistance and scaling challenge, are discussed in detail in section 4.

Finally, RMG cut is carried out to isolate different devices. In the proposed sequential CFET flow, it is carried out independently for top and bottom devices, which implies a reduced RMG cut etch depth. At a fixed maximum etch aspect ratio, a reduced etch depth has the benefit of a reduced RMG cut CD which defines gate line tip-to-tip (t2t) distance. This in turn allows to increase the nanosheet channel width for the same cell height, providing device performance improvement. Two potential drawbacks of an independent RMG cut are the additional overlay error introduced by separate top and bottom RMG cut lithography, and the requirement of an additional expensive high-numerical aperture (high-NA) gate cut mask, adding to the overall cost. Therefore, a common RMG cut process with a single mask may be considered for a more relaxed gate t2t requirement. Both cases (common and independent RMG cut) will be considered in the PPA assessment in section 5.

### 3. State-of-the-Art: Progress in 2D CFETs

Experimental 2D FET demonstrators are advancing rapidly, but they remain far from the process flow outlined in Section 2. Fig. 3a summarizes the increasing architectural complexity of these devices, starting from single-gate (SG) $WS_2$ FETs fabricated in the imec fab, to dual-gate (DG) structures with improved electrostatic control[34]. The third TEM in Fig. 3a shows a more advanced 2D GAA configuration, where the high-κ dielectric and metal gate fully wrap around the monolayer channel[35]. Finally, vertically stacked 2D nanosheets fabricated with a single gate stack deposition step have also been demonstrated[33]. This structure represents an important device-level precursor to the CFET integration flow outlined in Section 2.

Although the progression from planar SG to DG to GAA to stacked-GAA closely resembles the historical evolution of Silicon transistors, this evolution is not expected to deliver similar performance benefits in 2D FETs. While the change from SG to DG improves both the electrostatic control and the drive current, the change to GAA is not expected to further increase the on-current or improve the electrostatic control due to the monolayer thickness of the channel. Instead, the change from DG to GAA is driven by the integration flow outlined in Section 2, where the gate module is applied to all nanosheets simultaneously.

Experimentally, the added integration complexity of GAA 2D FETs comes with a performance gap compared to state-of-the-art SG or DG planar $MoS_2$ and $WSe_2$ lab devices. This is evident from the electrical benchmarking in Fig. 3b, where the 2D GAA FETs[17,36–40] show a penalty in SS and drive current compared to planar lab devices[41,42]. The benchmark considers various $L_{ch}$ = 20–100 nm, and the drive is normalized by the effective width ($W_{eff}$), as illustrated in Fig. 3c, to ensure a fair comparison of the different gate geometries. Importantly, this penalty is not inherent to the GAA configuration, but arises from the more challenging gate stack formation and the typically higher contact resistance in experimentally realized GAA devices. Additionally, the atomically thin channels are highly sensitive to metal deposition, stress, and process-induced damage.

A similar trend is observed for planar 2D FETs fabricated in the imec fab[34,43,44], which also underperform compared to lab-scale planar 2D FETs, despite benefitting from improved process control in the fab. This is largely due to tighter restrictions on available materials and processes in the fab. In particular, Titanium, Tungsten or Ruthenium have been used as contact materials, but these are largely inherited from silicon device processing, rather than optimized specifically for 2D FETs. In addition, processes commonly used for lab device processing, like contact lift-off, are not compatible with fab processing. As a result, alternative contacting schemes have been developed, like side contacts[34], top contacts by selective oxide etch[45], or bottom contacts planarized with CMP[43], all of which have a lower level of maturity. A second experimental performance gap is observed between 2D planar devices and state-of-the-art silicon GAA nanosheet FETs from imec[28] (Fig. 3b), primarily due to degraded subthreshold swing (SS) in 2D FETs at current levels above 10 nA/µm.Many reported planar 2D FETs show a similar stretch-out of the transfer curve, commonly attributed to band edge fluctuations, interface traps, Schottky contacts, or a combination of these effects. Notable exceptions are the InSe and $MoS_2$ channel devices with Yttrium contacts, linked to more Ohmic contact behavior[42,46], but these achievements have not been replicated so far. Even so, it is important to realize that monolayer 2D FETs are expected to outperform Silicon channels only at very short $L_g$, smaller than about 20 nm, due to their better electrostatic integrity (Fig. 1 b and c). At the longer $L_g$ considered in this benchmark, the 2D devices would at best match the highly optimized Silicon FETs.

While Fig 3b points out the yawning gap between demonstrated results from foundry Si GAA-FET and 2D-FETs, with the difference amplified as we go from LAB-based planar-FETs to LAB-based GAA-FETS to FAB-based planar-FETs, it is important to note that the rift can grow larger due to the integration choices associated with realizing the process-flow shown in Fig 2. Although these deleterious effects are yet unquantified, an appreciation of the probable deterioration and the sheer complexity of implementation, can be made by looking closer at some of the key process modules that have already been explored in 2D GAA FET integration. The first module (previously introduced in Fig. 2a) is the formation of the initial stack, consisting of alternating 2D channels, protection layers, and sacrificial (SAC) layers. The experimental demonstration in Fig. 3d uses $MoS_2$ channels surrounded by thin $Al_2O_3$ protection layers. These sacrificial protection layers provide mechanical stability and protection from damage later in the flow, and must ultimately be removable without harming the 2D sheet. Selecting the appropriate protection and SAC layers is critical, as these choices directly impact subsequent channel release, contact formation, and interface quality. In particular, oxides deposited by plasma-enhanced ALD (PE-ALD), plasma-enhanced chemical vapor deposition (PECVD), or sputtering often degrade the intrinsic properties of 2D materials because of energetic particle bombardment, plasma damage, or excessive defect generation. To mitigate these effects, more benign approaches such as thermal ALD or thermal evaporation are preferred. Conversely, unsuitable deposition methods or sacrificial-layer selections can lead to moisture uptake or stress accumulation within the 2D/SAC stack, which in turn promote interfacial instability and may result in delamination during subsequent processing.

The second key module (introduced in Fig. 2b) is the nanosheet etch, and realized by anisotropic etching step through the shallow trench isolation (STI) and 2D/SAC stack, creating a narrow vertical trench as shown in Fig. 3e. Maintaining a high-aspect-ratio profile with straight sidewalls in such a scaled structure is challenging because of the weak van der Waals interfaces; accordingly, a SiN hard mask is employed. A third key module (Fig. 2k) is the S/D contact formation, experimentally realized in Fig. 3f in two steps. The first is the formation of 2D protrusions by a controlled lateral recess of the SAC layers in the S/D regions. The selectivity of this recess etch is paramount: it must preferentially etch away the SAC, support layers or spacer material while leaving the 2D channels undamaged. The second step is the deposition of a contact material into the contact trench. Because the exposed 2D channels are located in deep narrow trenches, a conformal ALD process is needed to ensure that the metal wraps around the 2D sheets uniformly. The TEM in Fig. 3f showcases a thin TiN layer uniformly deposited by ALD along the trench sidewalls and making direct contact with the 2D channel edge. This contact scheme satisfies both conformality and material-integrity requirements: ALD enables gentle, uniform metal deposition without physical bombardment[47], while TiN offers a thermally stable interface to the 2D channel.

The channel release (Fig. 2h) is another critical module in the integration of 2D GAA FETs. Unlike silicon nanosheets, atomically thin 2D channels lack intrinsic mechanical rigidity, rendering them highly susceptible to collapse, distortion, or delamination during sacrificial-layer removal[48]. To maintain structural integrity throughout this step, further optimization of spacer and anchor designs is essential, such that the suspended 2D layers remain adequately supported while enabling complete and uniform release. In wet-etch-based approaches, the introduction of critical point drying is necessary to mitigate capillary forces during liquid–vapor transition, which otherwise pose a severe risk to freestanding monolayer channels. Whether using wet or dry release, the process must selectively remove the supporting material without damaging the 2D channel or undercutting adjacent dielectrics. Across both schemes, the use of support layers is beneficial when additional safeguarding of the 2D channel is required, as they can provide additional chemical shielding and mechanical reinforcement during etching and drying steps.

The RMG formation (Fig. 2i) is experimentally showcased in Fig. 3g, where a stack of $Al_2O_3$, $HfO_2$ and TiN conformally wraps around the 2D channel in a GAA configuration. Fig. 3g also illustrates ongoing refinements aimed at scaling the gate dielectric thickness from 10 nm to 5 nm while preserving interface integrity. A key enabler in this effort is the seeding layer, which facilitates uniform high-κ ALD on chemically inert 2D surfaces; however, it typically introduces additional EOT and exhibits a higher etch rate than the protection and sacrificial layers, necessitating careful selectivity control during subsequent processing. Although these results establish the feasibility of key 2D GAA process modules, the demonstrated devices in Fig. 3b still remain well short of the target specifications because of combined process-integration and contact-related limitations. This experimental gap motivates a broader comparison between 2D and Si CFET integration.

Compared with conventional Si-based CFETs, 2D CFETs differ substantially across several manufacturing modules (Table 1), and these differences help explain why the experimentally demonstrated building blocks have not yet translated into target-level device performance. First, wafer-scale formation of a 2D device stack requires either layer transfer or direct growth by techniques such as metal–organic chemical vapor deposition (MOCVD), whereas the remaining layers, including the sacrificial (SAC) dielectric and protection layers, can be deposited directly. Because adjacent layers in the 2D stack are coupled only through weak van der Waals interactions, adhesion is poor and the stack is therefore more vulnerable to delamination, particularly during high-shear processes such as chemical mechanical planarization (CMP). By contrast, Si CFETs rely on epitaxial stacking of Si and SiGe, which offers stronger adhesion and better processability. Second, the absence of dangling bonds on self-passivated 2D surfaces makes the uniform deposition of high-κ gate dielectrics and the control of their interfaces substantially more challenging. Third, monolayer 2D channels cannot tolerate conventional ion-implantation doping and are highly sensitive to process-induced stress, precluding the strain-engineering approaches widely used in Si CMOS.
Channel release process is also challenging on thin channels, as they are exposed to isotropic etch chemistry during the isotropic etch of SAC, as shown in Fig. 4c-c' for both Si and 2D GAA. In case of Si GAA, a ~5 nm thick Si film remains after SAC removal, which could be subsequently thinned to ~3 nm for a thin-channel device. For 2D-GAA on the other hand, the removal of SAC only leaves a 2D layer sandwiched between two protection layers for a total thickness of ~3 nm, which may be more susceptible to rupture during channel release and/or subsequent RMG high-k/metal gate deposition process. Additionally, in Si-GAA, S/D epi (Fig. 4b) can be used for channel stress engineering[49] which can be used to mechanically anchor the channel during channel release process (Fig. 4c). This is not the case in 2D-GAA, in which no S/D epitaxy is carried out. Instead, the ILD0 formation is carried out before channel release (Fig. 4a'-b'). Hence, the channel is only supported by inner spacers on two sides during channel release.

**4. Scaling limits of 2D CFET versus Si CFET**

Based on the proposed process flow in section 2, we now estimate the minimum achievable CPP with 2D GAA and compare it with that of Si GAA. As introduced in section 2, in Si GAA, to achieve low contact resistance, it is possible to grow S/D epitaxy (Fig. 4b). As shown in Fig. 2g, a minimum Contact width of 16 nm is assumed beyond which epitaxial growth may become challenging and contact resistance starts to degrade device performance[50]. On the other hand, in 2D-GAA, S/D epitaxy process is difficult on monolayer 2D sheets[51]. Therefore, S/D contacts may only be formed by direct, conformal metal deposition (Fig. 2f') with a suitable work function to overcome 2D-metal Schottky barrier to produce low-resistance ohmic contact. Achieving good contact between metal and 2D sheets also relies on the length of 2D sheet protrusion into the S/D region – wherein a zero protrusion would produce an edge-only contact, and a positive protrusion would produce a C-shape contact which includes both an edge as well as a surface contact with 2D sheet[48]. This is demonstrated in Fig. 4e' which further elaborates the S/D contact process in 2D GAA introduced earlier in Fig. 2k. First, contact trench is etched into ILD0 which exposes the edge of 2D channel (Fig. 4d'). A minimum CD of 10 nm is assumed to be able to etch an aspect ratio 4-5 contact trench into ILD0. This assumption is derived from the experimental demonstration of CFET contact trenches at CD~20 nm, AR~5 described in [52,53]. Then the inner spacer as well as the protection layer are isotopically recessed to create sheet protrusion into S/D region (Fig. 4e'). Compared to a non-protruded sheet, such a protruded 2D sheet provides a larger contact area and therefore, could make it easier to dope it to achieve lower contact resistance, as well as provides process margin to fabricate reproducible and reliable contact with subsequently deposited metal. A minimum of 3 nm sheet protrusion is assumed resulting in a 3 nm-wide protruded channel– metal contact (Fig. 4g). Hence, the total contact width in 2D GAA becomes 10 nm + 2*3 nm = 16 nm. For both schemes, we assume a minimum gate length of 10 nm (with 8 nm gate metal and 2 nm total high-k thickness on the sidewalls) and 5 nm inner spacer width.

Therefore, as shown in Fig. 4g, a minimum CPP of 36 nm is estimated for both Si and 2D-GAA architecture. Further scaling of contact width, Lg, spacer width and therefore, CPP can be argued for both the cases in equal measure indicating no CPP gains with 2D-GAA over Si-GAA.

**5. PPA challenges for GAA architecture based 2D CFETs**

Having established a feasible process flow for 2D CFETs, we now evaluate their circuit-level PPA to see if they can meet the A2 iso-frequency scaling target. The A2 node is where 2D CFETs could realistically be introduced, demanding extremely tight standard-cell dimensions but still requiring the same ring-oscillator frequency as the N2 node with only a modest (~8% per generation) increase in power density to compensate for the smaller area. According to the roadmap (Fig. 1), we are exploring a 2D CFET at the A2 node with a poly pitch (CPP) of 36 nm and a cell height (CH) of 42 nm, corresponding to an area of ~1500 nm².

We study a baseline A2 CFET design and three DoE variants (Table 2) that implement progressive architectural improvements. The baseline (DoE1) uses a common RMG cut shared by the top and bottom transistors, which restricts the total channel width to 7 nm within a 42 nm cell height (due to the 14 nm gate cut). DoE2 introduces separate RMG cuts for the nFET and pFET, a strategy discussed earlier to free up channel area. This reduces the gate-cut spacing from 14 nm to 8 nm, effectively doubling the combined channel width to 13 nm. DoE3 adopts an outer-wall forksheet (FS) structure – a technique already employed in advanced Si nanosheet nodes to gain effective width and reduce parasitics, which reduces the gate extension at the forksheet wall from 9 nm to 5 nm, increasing the channel width to 17 nm[54]. Finally, DoE4 explores idealized side contacts, permitting a tighter CPP (~30 nm); under iso-area conditions, this enables a cell height of ~50 nm and yields the largest effective width (25 nm). Note that such idealized edge-only contacts without any sheet protrusion into S/D, may not be process-viable as described in section 4. However, their assessment is important to understand the limiting case of maximum achievable effective width of 25 nm. All designs are evaluated at matched off-current ($I_{OFF}$) in an ~1500 nm² cell. As emphasized earlier, contact resistance ($R_c$) is a crucial factor for 2D devices; we therefore evaluate each design across a range of $R_c$ values (from an ideal 0 Ω·μm up to 100 Ω·μm) to gauge the impact of contact limitations on PPA. The list also includes $R_c$ = 42 Ω·μm, which is considered to be the state-of-the-art $R_c$ value for 2D-based devices in the literature[55].

It's worth noting that simply widening a device's channel would increase its drive current and channel capacitance approximately in tandem. If the parasitic capacitances also grew proportionally, the $I_{eff}/C_{eff}$ ratio would remain constant, and the design's point would slide along the same iso-frequency contour. Our design variants, however, aim to reduce the parasitic fraction of the total capacitance as width increases, enabling each successive DoE to achieve a higher $I_{eff}/C_{eff}$ ratio (i.e., to "jump" to a faster iso-frequency line in the $I_{eff}$–$C_{eff}$ space).

To perform the PPA analysis, we use a multi-scale modeling flow from materials to circuits. Ab initio simulations based on density functional theory (DFT) first provide the band structure and electron–phonon coupling for monolayer $MoS_2$ (nFET) and $WSe_2$ (pFET)[56,57]. These feed into non-equilibrium Green's function (NEGF) device simulations using ATOMOS[58], to produce intrinsic I–V and C–V curves for 10 nm gate-length devices under phonon-limited transport. We then fit a BSIM-CMG compact model to these intrinsic device characteristics (without any contact resistance or other extrinsic effects). Next, we build a detailed A2-standard-cell layout (INVD1) for each design and perform layout-aware parasitic extraction using Synopsys Raphael FX[59] . We incorporate the specified contact resistance Rc explicitly at the netlist level (as small series resistors at source/drain connections) and combine the compact device model with the extracted parasitics. Finally, we place each configured device in a 15-stage fan-out-of-3 (FO3) ring oscillator (no higher-level BEOL load) at $V_{DD}$ = 0.7 V to measure its effective switching current ($I_{eff}$) and effective switching capacitance ($C_{eff}$) for PPA benchmarking.

Figure 5c shows the measured $I_{eff}$ versus $C_{eff}$ for all four designs under varying $R_c$, with both axes normalized to the A2 target (red star at $I_{eff}$ = 1, $C_{eff}$ = 1). Higher $I_{eff}$ means more current drawn during switching (increasing speed potential but also dynamic power), while higher $C_{eff}$ means more charge needed per transition (slowing the device and raising energy per switch). Diagonal gray dashed lines from the origin indicate iso-frequency contours: points along each line have identical $I_{eff}/C_{eff}$ ratios and thus the same oscillation frequency. The A2 target lies on the 1.0× contour (orange dashed lines in Fig. 5c), meaning its $I_{eff}$ and $C_{eff}$ values yield the required N2-equivalent speed. Points above and left of the target's line (toward the green shaded "ideal scaling" region) can achieve equal or higher speed without extra power, whereas points below or to the right have lower $I_{eff}/C_{eff}$ ratios and cannot reach the target frequency unless supply voltage or other factors are changed.

Our design results in Fig. 5c reveal a clear trend. The baseline CFET (DoE1, $W_{ch}$ = 7 nm) sits at only ∼0.3–0.4× the target $I_{eff}$ and ∼0.7× the target $C_{eff}$ at $R_c$ = 42 Ω·μm, indicating a significant (> 50%) frequency shortfall relative to A2. For DoE2 ($W_{ch}$ = 13 nm), the separate gate cuts roughly double the effective width, leading to a sharp increase in $I_{eff}$ (now ∼0.7–0.8×) while adding only a little extra $C_{eff}$ (∼0.8×). This moves the DoE2 point much closer to the A2 iso-frequency line, boosting its frequency to ~70–80% of target at $R_c$

= 42 Ω·μm. DoE3 (FS, $W_{ch}$ ≈ 17 nm) further raises $I_{eff}$ (to ~0.8–0.9× at $R_c$ = 42 Ω·μm) with $C_{eff}$ around ~1.0–1.1×. The net $I_{eff}/C_{eff}$ ratio for DoE3 is high enough that, at very good contacts ($R_c \leq 10$ Ω·μm), it essentially meets the A2 speed. DoE4 (FS + side contacts, $W_{ch}$ = 25 nm) provides the highest intrinsic drive ($I_{eff}$ ~1.0× at $R_c$ = 42 Ω·μm), pushing that design's point near the A2 target. Under ideal contacts ($R_c$ = 0), DoE4 hits the target frequency (its symbol sits on the 1.0× line). At $R_c$ = 42 Ω·μm, DoE4 achieves ~90% of target speed, falling short of the full target performance.

To understand the role of capacitances, we examine the capacitance breakdown for each design in Fig. 5d. Each cluster of bars shows the absolute value (in fF) of the intrinsic channel capacitance (green), the metal-gate parasitic (teal), the gate via + input pin parasitic (blue), and their sum (orange). As $W_{ch}$ increases from 7 nm to 25 nm, the total measured capacitance (orange bar) roughly doubles, from ~0.05 fF in the baseline to ~0.09 fF in DoE4. Crucially, this increase is driven almost entirely by parasitic contributions (blue + teal bars); the channel capacitance grows proportionally with width and remains smaller than the parasitic contributions across all designs; however, its relative share increases in the widest variant (DoE4) as the parasitic components change only weakly with width.

The capacitance efficiency improvement is quantified in Fig. 5e, where we normalize each component by its channel capacitance ($C_{channel}$). Here the total (orange) bar indicates the capacitance overhead factor – i.e., how many times larger the total capacitance is compared to the channel's own capacitance. For the baseline ($W_{ch}$ = 7 nm), total capacitance is roughly six times the channel capacitance, indicating a heavy parasitic burden. With each successive DoE, this overhead factor falls: in DoE2 total capacitance is ~4× $C_{channel}$, in DoE3 ~3×, and in DoE4 ~2–2.5×. In short, the parasitic-to-channel capacitance ratio decreases monotonically across the designs, which confirms that our architectural changes are making each additional nanometer of width more effective. This directly correlates with the upward movement of the design points in Fig. 5c: a lower parasitic burden per width means a higher fraction of the added width goes into boosting $I_{eff}$ without proportionally dragging up $C_{eff}$, improving the overall $I_{eff}/C_{eff}$ balance.

Even so, the benefits from these architectural optimizations are incremental. The net $I_{eff}/C_{eff}$ ratio does improve with each DoE, but the baseline gap was sufficiently large that even our most advanced variant (DoE4) can only reach the A2 target speed under extremely low $R_c$ conditions, with essentially no margin. At a realistic $R_c$ = 42 Ω·μm, a gap remains. These results highlight that the main obstacles for monolayer 2D CFETs at A2 lie in parasitic capacitances and contact resistance rather than any fundamental limit of the 2D channels. In other words, the 2D transistors themselves offer strong intrinsic performance, but the tight CFET layout and associated parasitic capacitances increasingly dominate the total switching load, while the absence of doped contacts leads to high $R_c$ that limits current.

**Conclusions and Outlook**

2D semiconductors hold clear appeal for the A2 logic node because of their atomically thin channels and potential for extremely compact CFET designs. In this perspective, we have explored both the integration feasibility of monolayer 2D CFETs and their circuit-level performance limits against aggressive A2 PPA targets. Our analysis suggests that GAA-based CFET architectures face fundamental scaling limits beyond the A2 node, and that these limits are not specific to any one channel material but apply broadly to ultrathin-channel devices. In particular, once gate length and spacer dimensions approach their practical minima, contact formation and channel-release constraints set a similar lower bound on contacted poly pitch, such that the electrostatic advantages of thinner channels no longer translate into continued layout scaling.

Although 2D FETs can be integrated in principle within a GAA CFET architecture at a contacted poly pitch of 36 nm at A2 node, several practical constraints blunt their benefits. The most notable issue is high contact resistance ($R_c$), a consequence of the difficulty of forming low-resistance ohmic contacts with ultra-thin 2D channels in the absence of heavy doping. We also found that the tight A2 layout introduces significant parasitic capacitances (from gate and local wiring structures) which, unlike intrinsic channel capacitance, do not scale down with decreasing device width. These factors together lead to substantial performance loss: the baseline 2D GAA CFET design in our analysis falls more than 50% short of the A2 target frequency, illustrating the magnitude of today's gap.

We addressed part of this gap by exploring architectural optimizations (DoE1–DoE4) to recover channel width and improve capacitance efficiency. The results indicate that each successive improvement can raise $I_{eff}/C_{eff}$ and shrink the performance deficit. However, even under aggressive design assumptions and maximum active sheet width, none of the designs meet the A2 frequency target due to fixed parasitic capacitance overheads, unless extremely optimistic contact resistance is assumed. .

These findings suggest that 2D CFETs will not naturally achieve their full promise by simply being inserted into a silicon-optimized GAA architecture. This is not due to a fundamental flaw in 2D channels, but rather to a mismatch between the conventional CFET design and the unique properties of ultra-thin 2D materials. To bridge that gap, co-optimization across multiple fronts is needed. First, contact resistance must be dramatically reduced. Continued efforts in 2D contact engineering—such as improved metal–2D interfaces or alternative contact schemes—are essential for unlocking higher drive currents and mitigating the $R_c$ bottleneck. Second, future 2D CFET designs must reduce parasitic capacitances in order to meet the A2 performance target.

Finally, to fully exploit 2D materials, the logic roadmap must pivot away from a strict reliance on classical GAA CFET concepts and explore new 2D-specific CFET architectures that relax some of the constraints intrinsic to the Si GAA CFET structure. Promising directions include hybrid planar–vertical CFET schemes that trade full gate enclosure for reduced parasitics. These shared-gate or asymmetric-gate configurations lower architectural overhead and enable designs that intentionally decouple electrostatic control from contact formation. While no single architecture has yet emerged as the definitive successor to GAA, the message is clear: the next gains will come from rethinking the transistor architecture itself, not from further optimization of a silicon-era template. If such co-optimization can be achieved, 2D CFETs remain a credible—and potentially indispensable—option for extending Moore's law into the A2 and post-A2 era, where energy efficiency and density, rather than brute-force electrostatics, will define the success of logic technologies.

**Acknowledgments**

We acknowledge the contributions of the imec pilot line, imec Xplore Lab, and AMSIMEC teams, in particular Bavo Storms, Rudy Verheyen, Han Han, and Olivier Richard, for their assistance. We also thank Halil Kuekner for layout discussions and Alex Makarov for support with the PPA software. The authors also thank Lam Research COVENTOR for access to the SEMulator3D software. The authors further thank the entire imec 2D project team for the discussions that preceded the preparation of this manuscript. This research was funded by the imec IIAP core CMOS programs.

**Author contributions**

F. X., G. G., and C. C. R. conceived the idea and initiated the project. F. X. and K. B. planned and designed the research framework. A. G., F. X., and J. B. designed the process flow. G. G., S. Y., and A. A. carried out the device and circuit simulations. F. X., K. B. and W. J. designed and completed the device fabrication. M. P. and G. H. provided input on the device and circuit simulations. F. X. X. W., T. C., J. M.

and Q. S. performed the device characterization and benchmarking. G. S. K., G. H., and C. C. R. supervised the project. All authors discussed the results and commented on the manuscript.

**Competing interests**

The authors declare no competing interests.

**References to be updated after review**

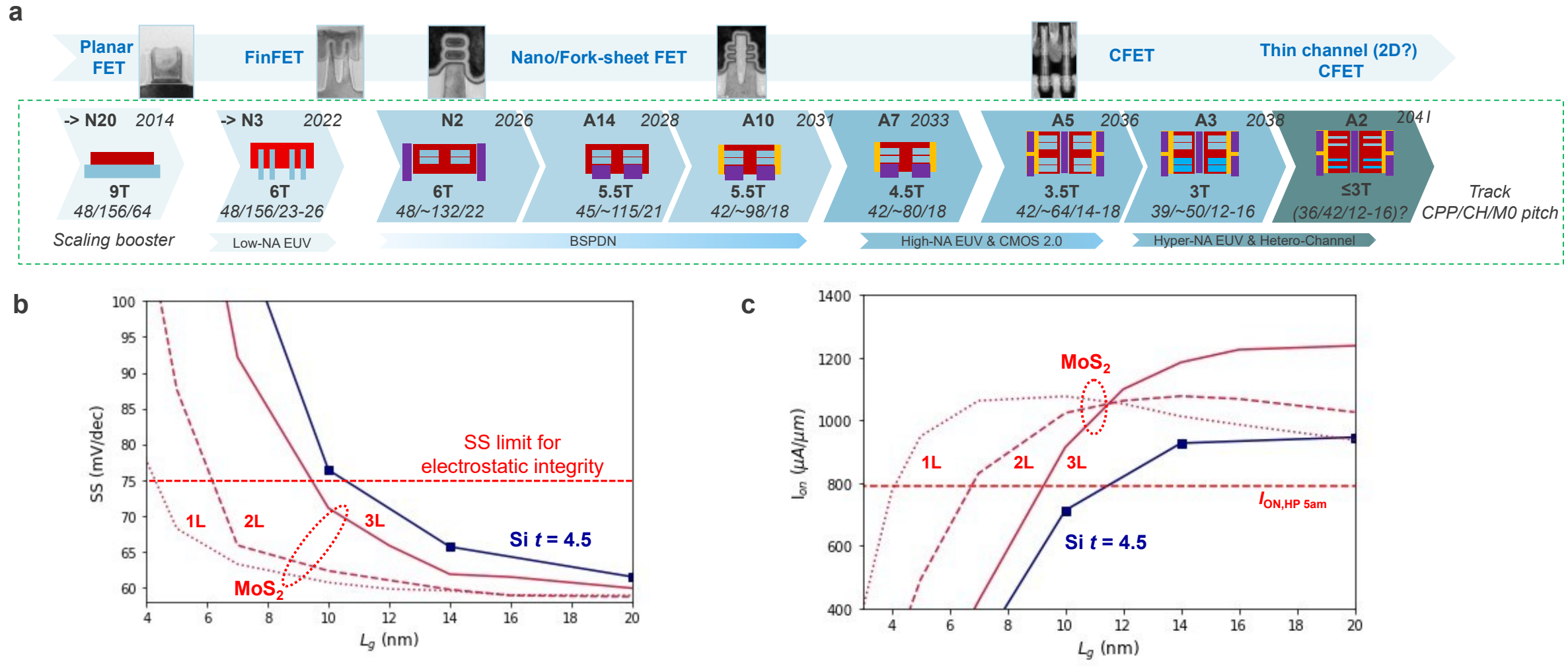

**Fig. 1 | Introducing 2D-material-based devices in the logic scaling roadmap. a,** imec Logic research roadmap for HPC[54,60–63]. CPP is the contacted poly pitch (minimum gate-to-gate spacing), CH is the standard cell height (vertical cell dimension), and M0 pitch is the pitch of the lowest metal interconnect layer used for local routing. A 6T standard cell refers to a cell whose height accommodates six M0 routing tracks. DFT–NEGF-simulated scaling potential of mono- (1L, $t \approx 0.7$ nm), bi- (2L, $t \approx 1.4$ nm) and tri- (3L, $t \approx 2.1$ nm) layer $MoS_2$ and $t = 4.5$ nm Si nanosheet (NS) FETs. **b**, $SS$ vs. $L_g$ and **c**, $I_{ON}$ vs. $L_g$ ($I_{OFF}$ = 5 nA/µm) for optimized Si and $MoS_2$ NS at $V_{DD}$ = 0.6V. The current is normalized per gate width. The simulations were performed with our quantum transport solver ATOMOS[64]. The methodology and models used for $MoS_2$ are fully described in [64], while those used for Si are detailed in [58,65]. Equivalent oxide thickness (EOT) = 0.5 nm.

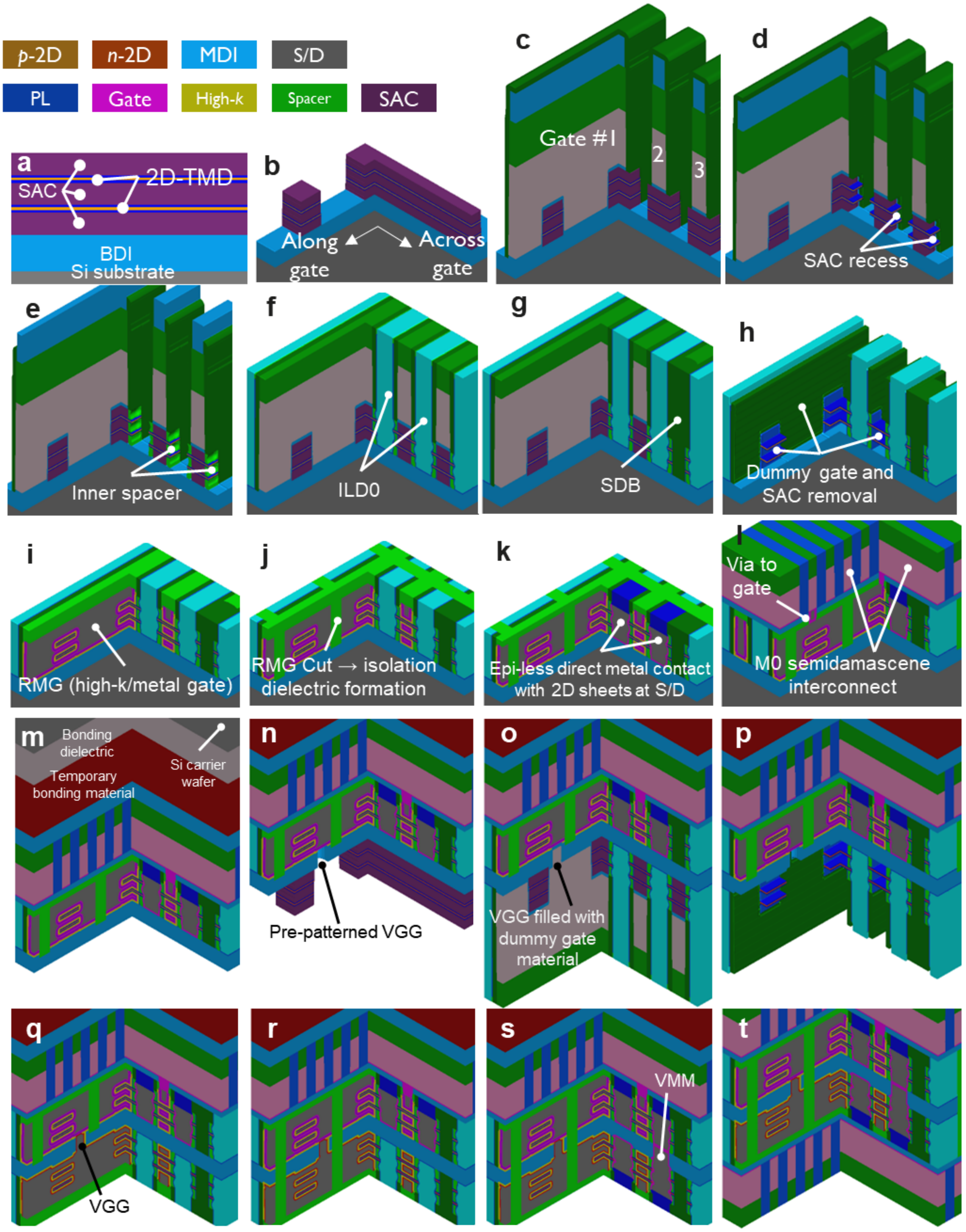

**Fig. 2 | Sequential CFET beyond A2 node process flow cross-sectional view along gate #1, and across gates#2 and 3 gate direction with 2D-TMD channels.** A bottom dielectric isolation (BDI) is used to aid sequential top and bottom device formation. **a,** Stack formation: BDI layer deposition, deposition or transfer of n-type 2D–TMD layers, deposition of sacrificial (SAC) layer and protection layer (PL). **b**, Top nanosheet etch. **c**, Top dummy gate etch and gate spacer deposition and etch, and exposed nanosheet stack etch. **d**, Sacrificial material recess (Recess depth = Inner spacer CD + Channel

protrusion). **e,** Inner spacer formation. **f,** Interlayer dielectric (ILD0) deposition, anneal and CMP. **g,** Single diffusion break (SDB) which renders gate #3 non-functional for physical isolation between adjacent gates. **h,** n-2D channel release by dummy gate pull. **i,** 2D NMOS Replacement metal gate (RMG) formation which includes gate high-k dielectric and gate metal deposition, CMP recess and plug formation. **j,** 2D NMOS RMG Cut and isolation formation. **k,** 2D NMOS S/D formation (shown here with edge-only 2D sheet to metal contact). **l,** Via to 2D NMOS S/D, Via to Gate and first BEOL metal ($M_0$) formation & remaining front-side BEOL (not shown for brevity). **m,** Wafer bonding and wafer flip for backside processing + Si grinding and CMP for backside reveal. **n,** Bottom device stack formation + Bottom nanosheet etch and VGG pre-patterning. **o,** Similar to steps c-g for bottom device. **p,** p-2D channel release by dummy gate pull. **q,** 2D PMOS RMG formation including VGG metallization. **r,** 2D PMOS RMG Cut formation. **s,** 2D PMOS S/D contact formation with a top-bottom S/D connecting via (VMM). **t,** Via to 2D PMOS S/D, Via to Gate and BM0 formation & remaining back-side BEOL (not shown for brevity).

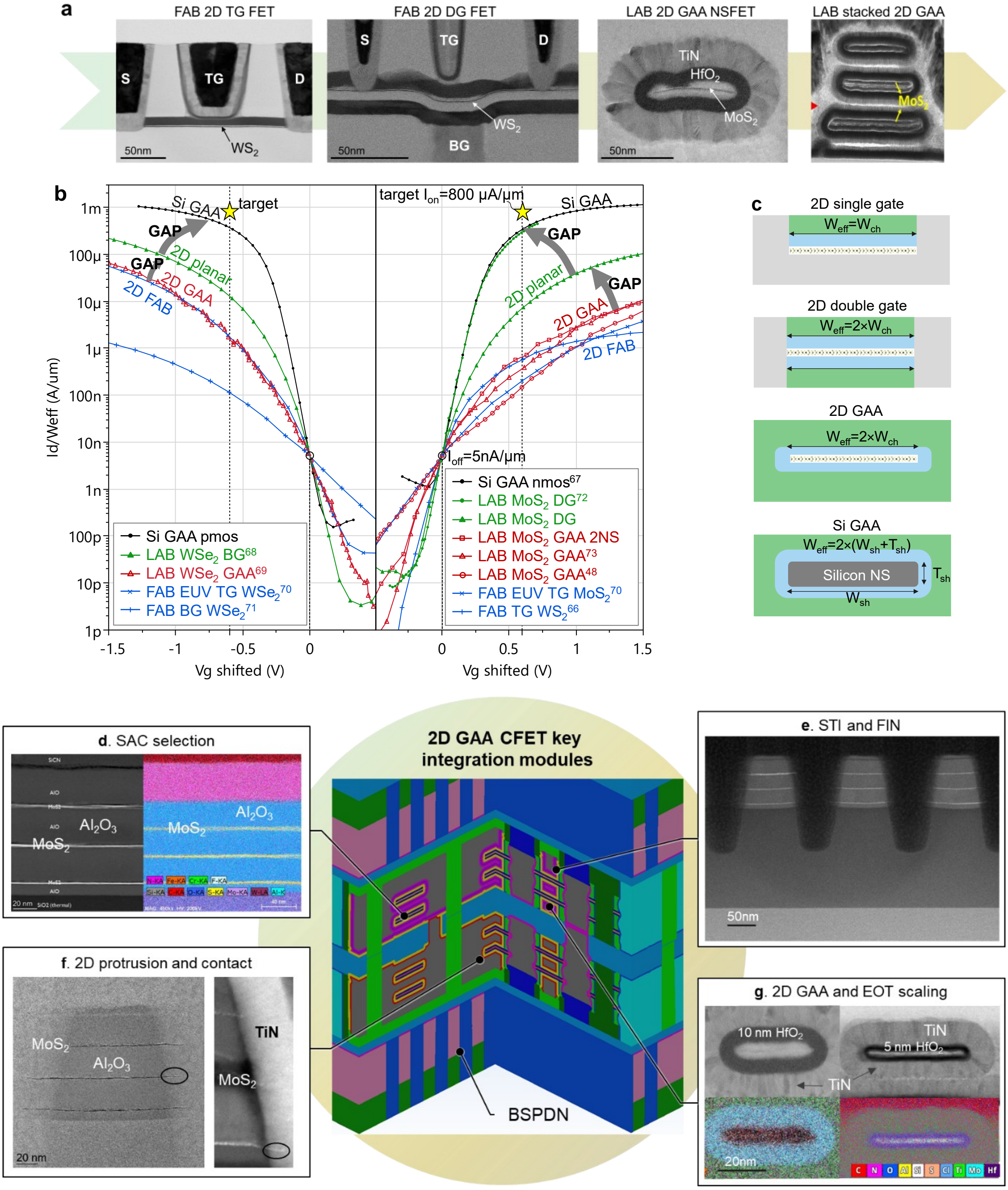

**Fig. 3 | Evolution of 2D FETs, benchmarking, and key modules demo and challenges for 2D nanosheet FETs/CFETs based on GAA architectures. a**, Increasing complexity of 2D FETs, from SG and DG[66], to GAA FETs[40] and stacked-GAA FETs [48]. b, Benchmarking of experimental Si GAA FETs from imec, 2D planar, 2D GAA and 2D fab devices. Transfer curves are shifted horizontally to enforce $I_{OFF}$ = 5 nA/μm, enabling a simple comparison of ION at fixed overdrive of 0.6 V, matching the target VDD. This benchmarking metric is more stringent than the typically reported Imax versus *I*min or versus SSmin. Si GAA pmos (previously unpublished) has $W_{sh}$ = 27 nm, $t_{sh}$ = 8 nm, $W_{eff}$ = 70 nm, *L*g~20 nm, $V_d$ = −0.7 V. Si GAA nmos [67] has $W_{sh}$ = 22 nm, $t_{sh}$ = 8 nm, $W_{eff}$ = 60 nm, $L_g$~20 nm, $V_d$ = 0.7 V. LAB $WSe_2$

BG [68] has $L_{ch}$=100 nm, $W_{ch}$=$W_{eff}$=1 µm, $V_d$=−1 V. LAB $WSe_2$ GAA [69] has $L_{ch}$=20-80 nm, $V_d$=−1 V. FAB EUV TG WSe2 [70] has $W_{ch}$ = $W_{eff}$ = 76 nm, $L_{ch}$ = 30 nm, $L_{cont}$ = 30 nm, EOT~2 nm, $V_d$=−1 V. FAB EUV TG MoS2 [70] has $W_{ch}$ = $W_{eff}$ = 646 nm, $L_{ch}$ = 28 nm, $L_{cont}$ = 22 nm, EOT~2 nm, $V_d$=1 V. FAB BG WSe2 [71] has $L_{ch}$=210 nm, $W_{ch}$=$W_{eff}$=1 µm, $V_d$=−1 V. LAB MoS2 DG (circles) [72] has $L_{ch}$ = 20 nm, $V_d$ = 0.7 V. LAB $MoS_2$ DG (triangles, previously unpublished) has $L_{ch}$ = 50 nm, $W_{ch}$ = 1 µm, $W_{eff}$ = 2 µm, $V_d$ = 1 V. LAB $MoS_2$ GAA 2NS (previously unpublished) $L_{ch}$=200 nm, $W_{ch}$ = 80 nm, $W_{eff}$ = 320 nm, $V_d$=1 V. LAB $MoS_2$ GAA [73] has $L_{ch}$ = 50 nm, $V_d$ = 1 V. LAB MoS2 GAA [48] has $L_{ch}$ = 40 nm, $W_{ch}$ = 10×50 nm, $W_{eff}$ = 1 µm, $V_d$ = 1 V. FAB TG $WS_2$ [66] has $L_{tg}$ = 100 nm, $L_{ch}$ = 240 nm, $V_d$ = 1 V. The drain current is normalized with the effective width depicted in c, streamlining comparison with Silicon FETs and forming a stepping stone to the PPA analysis in Section 4, where the increased capacitance of 2D GAA devices is taken into account and normalization with Weff is considered. d, Cross-sectional TEM images illustrating the stacked 2D channel and sacrificial-layer (2D/SAC) structure, e the etched fin geometry, f the selectively released 2D channel protrusion and S/D contact formation, and g the completed gate stack in a 2D GAA configuration.

| Integration challenge | Si GAA CFET A3 | 2D GAA CFET beyond A3 |
|---|---|---|
| Nanosheet formation | High-quality, selective epitaxial Si/SiGe multilayer formation, followed by monolithic etching of the Si and SiGe stack. Ultimate channel scaling is limited by Si-Ge interdiffusion. | 2D/SAC multilayer formation by direct growth or transfer. There is an increased risk of delamination, line collapse, or pillar collapse due to detachment at the interfaces with the van der Waals-bonded 2D layer. |
| Protrusion formation and source/drain contact | Highly doped lateral Si(C) or SiGe epitaxy on the Si sheet edges enlarges the contact area. Doping occurs through diffusion from the highly doped contacts. | Selective doping of the region under the spacer is challenging. The contact area can be increased by ALD or CVD wraparound contacts on the protruding 2D channel, but mechanical stability is required. |
| Channel release | Low mobility and reduced mechanical robustness at a Si channel thickness of ~3 nm during channel release (device-dimension dependent). | A 2D monolayer channel also has limited mechanical robustness, but it can benefit from interfacial layers for channel protection and improved mechanical robustness during channel release (device-dimension dependent). |
| Gate stack | High-k and gate metal are well developed, including controlled Si surface oxidation used as a seeding layer for high-k growth. | The high-k gate stack on 2D materials in replacement-gate mode needs further optimization to reduce defect density, and the seed layer used on 2D materials is more difficult to scale. |
| Integration approach | Monolithic, sequential, or hybrid integration; low-/high-NA EUV. | Monolithic, sequential, or hybrid integration; hyper-NA EUV or beyond. |

**Table 1.** Integration challenges of 2D CFETs compared with Si CFETs

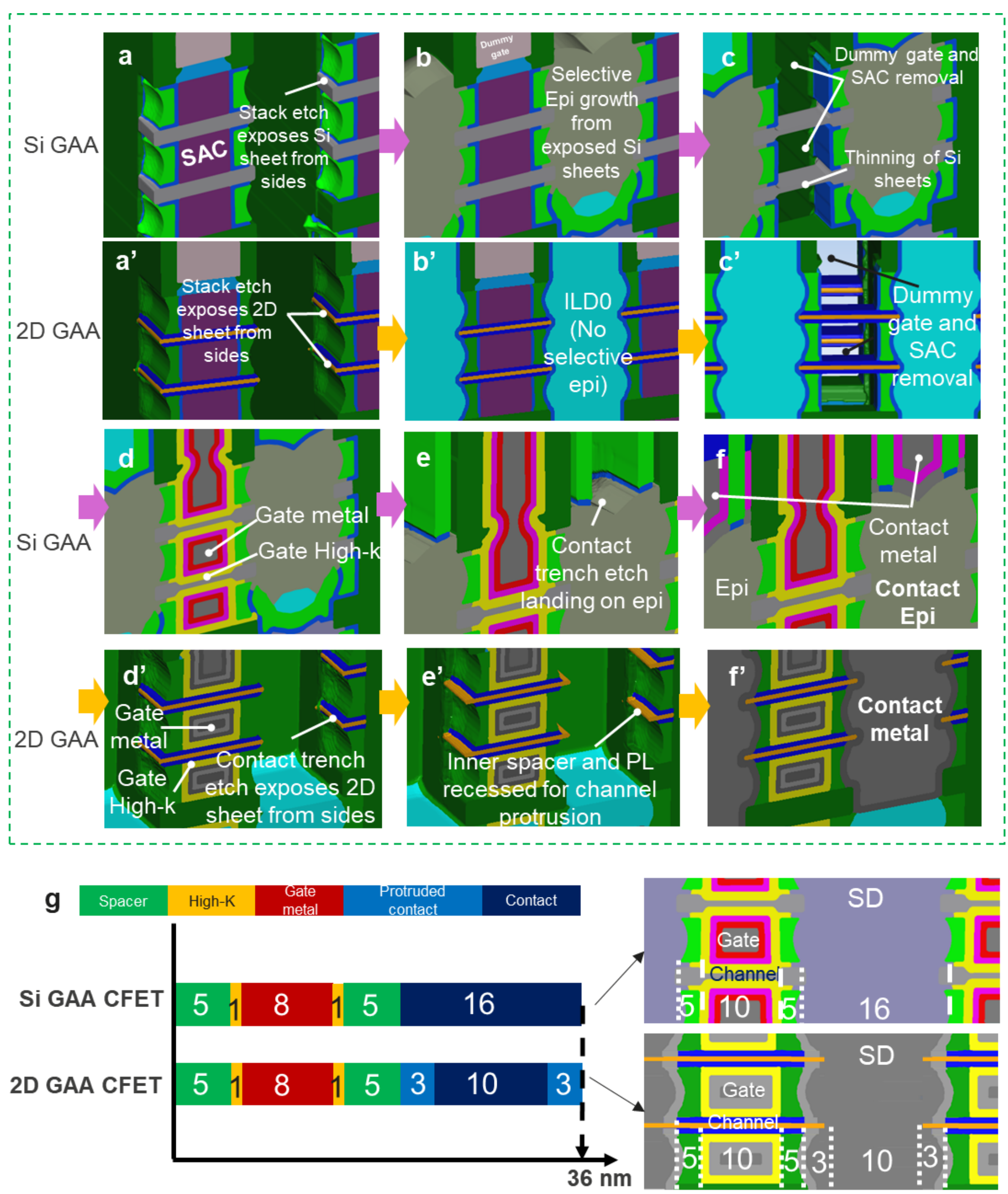

**Fig. 4 | Scaling limits of CFET with GAA architecture.** Challenges with thin channel release process in (**a-c**) Si GAA vs. (**a'-c'**) 2D GAA; Differences in S/D contact process between (**d-f**) Si GAA in which S/D epitaxy is followed by S/D contact trench etch and metallization vs. (**d'-f'**) 2D GAA in which the inner spacer and protection layer are recessed to create 2D sheet protrusion to achieve lower contact resistance and a reliable contact. Direct metallization is carried out on top of protruded 2D sheets. **g**, Break-down of minimum achievable Contact poly pitch (CPP) in Si GAA vs. 2D GAA showing a minimum CPP = 36 nm for both the cases.

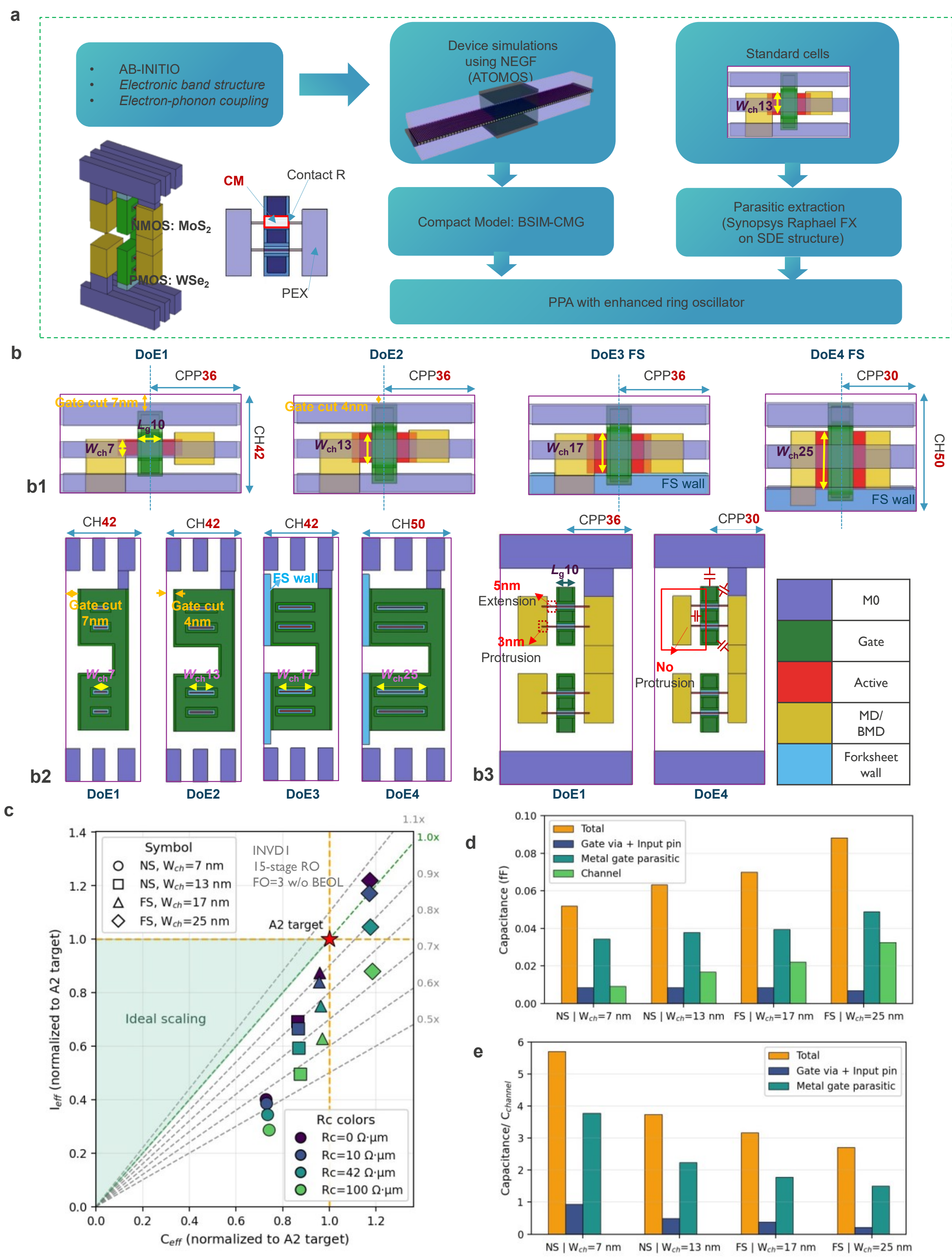

**Fig. 5 | Power–performance assessment of 2D GAA devices listed in Table 2. a**, DTCO framework for 2D GAA devices consists of two main parts: 1) **2D channel modeling**: ab initio DFT is used to extract the electronic band structure and electron–phonon coupling of the 2D material, and these outputs are input into ATOMOS to generate device I–V and C–V characteristics, followed by calibration of a BSIM-CMG compact model to match the ATOMOS results; 2) **Parasitics extraction**: A Sentaurus Structure Editor (SDE)[74]. structure of an inverter with drive strength 1 (INVD1) is constructed, and

Synopsys Raphael FX is used to extract parasitic resistance and capacitance values excluding the channel part. By combining the calibrated channel model with the extracted parasitics, a complete INVD1 netlist is obtained, which is then used in a 15-stage ring oscillator with FO3 without BEOL to evaluate device power and performance; **b1**: Top view of SDE structures for DoE1–4, illustrating the increase in active sheet width from 7 nm to 25 nm. **b2**: Cross-sectional view along the gate for DoE1–4, highlighting the techniques used to enhance active sheet width. **b3**: Cross-sectional view along the active region for DoE1 and DoE4; in DoE4, scaling the CPP to 30 nm prevents sheet protrusion into the S/D metal contact metal. **c**, RO results for four DoEs with four different $R_c$ values. $I_{eff}$, average current measured from the $V_{DD}$ rail during switching, is plotted against $C_{eff}$. $C_{eff}$ is defined by $I_{eff}$/frequency/$V_{DD}$/ln(2). Power is $V_{DD} \times I_{eff}$. Results are normalized to the A2 target, defined to have iso-frequency relative to N2 nanosheet, with current scaled to an 8% increase in power density per node. The results show that increasing the channel width enhances both the drive current and the oscillation frequency. The device with the largest channel width of 25 nm meets the frequency target only when the contact resistivity is ≤ 10 Ω·μm. **d**, Capacitance breakdown for the four DoEs, including contributions from parasitic components (M0 input pin, gate via, and metal gate) and the channel. The results show that parasitic capacitances account for a significant fraction of the total capacitance . **e**, Ratio of total and parasitic capacitances to channel capacitance for the four DoEs. By normalizing to channel capacitance (which determines the drive current), the results show that optimized device architectures and increased channel width effectively reduce non-channel capacitances, leading to improved frequency performance.

| | CPP [nm] | CH [nm] | Wch [nm] | RMG cut of top/bottom device | RMG Cut CD | Type | Gate extension length [nm] | Sheet protrusion length into S/D [nm] |
|---|---|---|---|---|---|---|---|---|
| DoE1 | 36 | 42 | **7** | Common | 14 | NS | 9 (no-VGG side), 12 (VGG side) | 3nm |
| DoE2 | 36 | 42 | **13** | **Independent** | **8** | NS | 9 (no-VGG side), 12 (VGG side) | 3nm |
| DoE3 | 36 | 42 | **17** | **Independent** | **8** | **FS** | 5 (FS wall side), 12 (VGG side) | 3nm |
| DoE4 | **30** | **50** | **25** | **Independent** | **8** | FS | 5 (FS wall side), 12 (VGG side) | **0nm** |

**Table 2. 2D GAA structure DoEs.** All devices have a gate length ($L_g$) of 10 nm, spacer thickness of 5 nm and EOT of 0.8 nm. DoE1 uses a single-sided RMG cut, resulting in a 14 nm gate cut and a 7 nm channel width. In DoE2, a double-sided RMG cut reduces the gate cut to 8 nm and increases the sheet width to 13 nm. DoE3 introduces a forksheet structure, reducing the gate extension at the wall side to 4 nm and further increasing the sheet width to 17 nm. In DoE4, the CPP is scaled to 30 nm; under iso-area conditions, the cell height increases to 50 nm, enabling a sheet width of 25 nm. However, at 30 nm CPP, gate-extension protrusion is no longer feasible.